\documentclass{aa}    
\usepackage{graphics}

\newcommand{\apj}{ApJ}
\newcommand{\mnras}{MNRAS}

\begin{document}

\title{Magnetic field evolution
       in galaxies interacting with the intracluster medium}
\subtitle{3D numerical simulations}
\author{K. Otmianowska-Mazur \inst{1}  
\and
B. Vollmer \inst{2}
}
\institute{Astronomical Observatory
	   Jagiellonian University
   Krak\'ow, Poland \email{otmian@oa.uj.edu.pl}
\and
Max-Planck-Institut f\"{u}r Radioastronomie, Auf demH\"{u}gel 69, 53121 Bonn, Germany} 

\date{Received May 28, 2002}

\abstract{
A fully three-dimensional (3D) magnetohydrodynamical
(MHD) model is applied to simulate the evolution of the large-scale 
magnetic field in cluster galaxies interacting with the intra-cluster 
medium (ICM). As the model input we use a time dependent
gas velocity field resulting from 3D N-body sticky-particle 
simulations of a galaxy.
The modeled clouds are affected by the ram pressure due to their rapid
motion through the ICM in the central part of a cluster. 
Numerical simulations have shown that after the initial compression phase due to 
ram pressure a process of gas re-accretion onto the galactic disk takes place.
We find that the gas re-accretion leads to an increase of 
the total magnetic energy without any  dynamo action.
The simulated magnetic fields are used to construct the model maps of  
high-frequency (Faraday rotation-free) polarized radio emission.
We show that the evolution of the polarized intensity shows features that are
characteristic for different evolutionary stages of an ICM--ISM interaction.
The comparison of polarized radio continuum emission maps with our model
permits to determine whether the galaxy is in the compression or in the
re-accretion phase. It also provides an important  constraint
upon the dynamical modeling of an ICM--ISM interactions.
}


\authorrunning{Otmianowska-Mazur \& Vollmer}
\titlerunning{Magnetic field evolution of galaxies}

\maketitle

\section{Introduction}  

It is well known that galaxies in clusters have different
physical and morphological  properties than field galaxies.
They are often H{\sc i} deficient, i.e. they have lost up to 90\% of their
atomic gas - mainly due to the interaction of the interstellar medium
(ISM) with the intra-cluster medium (ICM) (Solanes et al. 2001). 
The  cluster galaxies often possess H{\sc i} tails and strong compression 
regions (Cayatte et al. 1994, Bravo-Alfaro et al. 2000,
Chy\.zy et al. in prep.). 
These observed asymmetries of the gas distribution can be explained  
either by ram, pressure effects (Vollmer et al. 2001 and references
therein) and/or by tidal interactions with nearby companions 
(e.g. NGC~4438, Combes et al. 1988).
Recent radio continuum observations of Virgo Cluster galaxies
show also magnetic field structures  which are not 
present in normal galaxies (Soida et al. 1996, Vollmer et al. in prep.).
NGC~4254, oriented face-on, shows strong radio polarization maxima
distributed outside the optical spiral arms in the direction
towards the Virgo Cluster center (Soida et al. 1996, Chy\.zy et al., 2001).
The highly inclined galaxy NGC~4522 also shows asymmetric
radio polarization destinations (Vollmer et al. in prep.).

While the question of the gas reaction to the ICM ram pressure
has a rather long history (Takeda, Nulsen, \& Fabian 1984, 
Gaetz, Salpeter \& Shaviv 1987; Balsara, 
Livio, \& O'Dea 1994; SPH: Tosa 1994; Abadi, Moore, \& Bower 1999,
Schulz \& Struck 2001;  Vollmer et al. 2001), 
there were no investigations yet concerning the evolution of magnetic 
fields in cluster galaxies.
Therefore, a comprehensive study of magnetic field behaviour
under the influence of the gas stripping process is needed to provide a physical 
interpretation of the observed features in the radio continuum
and polarization in stripped spiral galaxies.

The evolution of the magnetic field structure in galaxies is
closely related to the amplification of the large scale magnetic field by
a dynamo action. The main clue of the traditional MHD dynamo is the rate
of small scale turbulent motions in the rotating medium  producing
large-scale  poloidal magnetic field  from the azimuthal one. Then the 
differential rotation reproduces the toroidal magnetic field with 
its strength amplification (e.g. Parker 1979 and references therein).

We solve the induction equation for the gas velocity fields obtained 
from the N--body simulations by Vollmer et al. (2001).
With our simulations we analyze the problem of a total magnetic energy 
amplification in galaxies interacting with the ICM.
In a first step no dynamo action is included.

The main goal of our project is to study a representative 
case of ICM--ISM 
interaction to investigate the characteristic features of the  magnetic field 
evolution.
This will be a first step to understand the observed magnetic fields in 
cluster galaxies.

The outline of this article is as follows: the model used to simulate the
evolution of the magnetic field is introduced in Sect.~\ref{sec:model}.
The resulting evolution of the magnetic field topology 
is shown in Sect.~\ref{sec:topology}.
In Sect.~\ref{sec:energy} the evolution of the total magnetic energy is 
investigated, followed by the construction of model maps of polarized radio
emission (Sect~\ref{sec:polmaps}). We discuss our results in 
Sect.~\ref{sec:discussion} and summarize them in Sect.~\ref{sec:conclusions}.

\section{Model \label{sec:model}}

We constructed a 3D MHD model of magnetic field
evolution (see Otmianowska-Mazur et al. 1997 for more details)
in a galaxy moving rapidly through the ICM.
We  apply the Zeus3D code  (Stone \& Norman 1992a and b) solving
the induction equation: 
\begin{equation}
{\partial\vec{B}/\partial t=\hbox{rot}(\vec{v}\times\vec{B})
 -\hbox{rot}(\eta~\hbox{rot}\vec{B})}
\label{eq:inductioneq}
\end{equation}
where $\vec{B}$ is the magnetic induction, $\vec{v}$ is the large-scale
velocity of the gas, and $\eta$ is the coefficient of a turbulent diffusion.
Realistic, time dependent gas velocity fields are
provided by the 3D N-body sticky-particle simulations of H{\sc i} cloud
complexes evolving in an analytically given gravitational potential 
of a disk, a bulge and a halo (Vollmer et
al. 2001). The clouds collide inelastically and are affected  
by the assumed ICM ram pressure due to relative motion of the galaxy
with respect to the ICM.

\subsection{Model input parameters}

We model a galaxy which is mildly H{\sc i} deficient  (it has lost
$\sim$30\% of its gas) at the end of the simulation. 
It is on an eccentric orbit within the cluster and its disk plane
is inclined by 20$^{\circ}$ with respect to the orbital plane.
When the galaxy is approaching the cluster center the clouds are pushed
and possibly stripped away in the direction opposite
to its velocity respect to ICM. When the galaxy emerges
from the cluster core, a process of cloud re-accretion starts due to 
the  galactic gravitational  potential.
The sticky-particle code uses 10000 cloud complexes with different
masses. Ram pressure is modeled as an additional acceleration on the 
clouds located at the surface of the gas distribution in the direction of 
the galaxy's motion within the cluster. Thus, there is no explicit 
intra-cluster medium included in the model.
The maximum ram pressure applied on the clouds corresponds
to an intra-cluster medium density 
of $2\times 10^{-3}~\rm cm^{-3}$ (see Vollmer et al. 2001 for more details)
and a maximum velocity of the modeled galaxy of about 1500~km\,s$^{-1}$
with respect to the cluster center. 
The time step of our calculations is $10^7$~yr. The time of closest
approach to the cluster center is $t$=500~Myr after the start of the
simulations.

The induction equation is analyzed in rectangular coordinates ($XYZ$).
The number of grid points used is 171x171x71 along the $X$, $Y$ and $Z$ axis, 
respectively. This corresponds to the 
grid spacing of 200~pc in the galactic plane and
of 280~pc in the $Z$ direction, resulting in a size of the modeled box of
34~kpc~x~34~kpc~x~20~kpc. Since the N-body code is discrete
whereas the MHD code is using a grid, we have to interpolate the
discrete velocities on the grid. This is done using a spline
function with a density dependent smoothing length. It turned out that we
had to use a large smoothing length  to suppress the noise 
in the velocity field of the outer disk, which is due to a small, local particle density.
In this way we avoid numerical artifacts of the magnetic field distribution
at the outer disk.
This has the consequence that the velocity field at the edge of the gas
distribution is more extended than the gas distribution itself,
which leads to an overstanding magnetic field at the edges of the gas distribution.

The large number of grid points gives rise to a computational 
difficulty: in our earlier papers (eg. Otmianowska-Mazur et al. 2002, 
Elstner et al. 2000)
the cell size in $Z$ direction was 80~pc, whereas in the present simulations
it is 280~pc, thus the value of the numerical diffusion in the $Z$ direction is 
certainly higher. In order to avoid such
diffusion in the central part of our box during the first 20 time steps
the velocity curve was smoothed from the distance of 5~kpc
to the center with a Gaussian function with $\sigma = 2$~kpc.
This smoothing influences the magnetic field
configuration in the later time steps only marginally, 
but allows to keep magnetic flux 
in the central disk constant during the first 200~Myr of its evolution.
We assume the magnetic field to be partially coupled to the gas via the 
turbulent diffusion process (Otmianowska-Mazur et al. 2002).
We made two basic experiments with two different behaviours of the
turbulent diffusion coefficient: uniform (case A) and 
linearly growing with $z$ in the halo (case B) 
(see Otmianowska-Mazur et al. 2002 for a detailed explanation). 
Its value is  $5\times10^{25} \rm cm^2 \rm s^{-1}$ in the whole box
for  case A.  For experiment B it has the same value in 
the disk plane.
To check how  numerical diffusion influences our results,
two additional computations were made:
no physical diffusion (only numerical diffusion) (case C),
and uniformly distributed diffusion 5 times smaller than in case A (case D).
The initial magnetic field is purely toroidal with a strength of 1~$\mu$G.

\section{Results \label{sec:results}}

\subsection{Magnetic field geometry \label{sec:topology}}

\begin{figure*} 
\resizebox{\hsize}{!}{\includegraphics{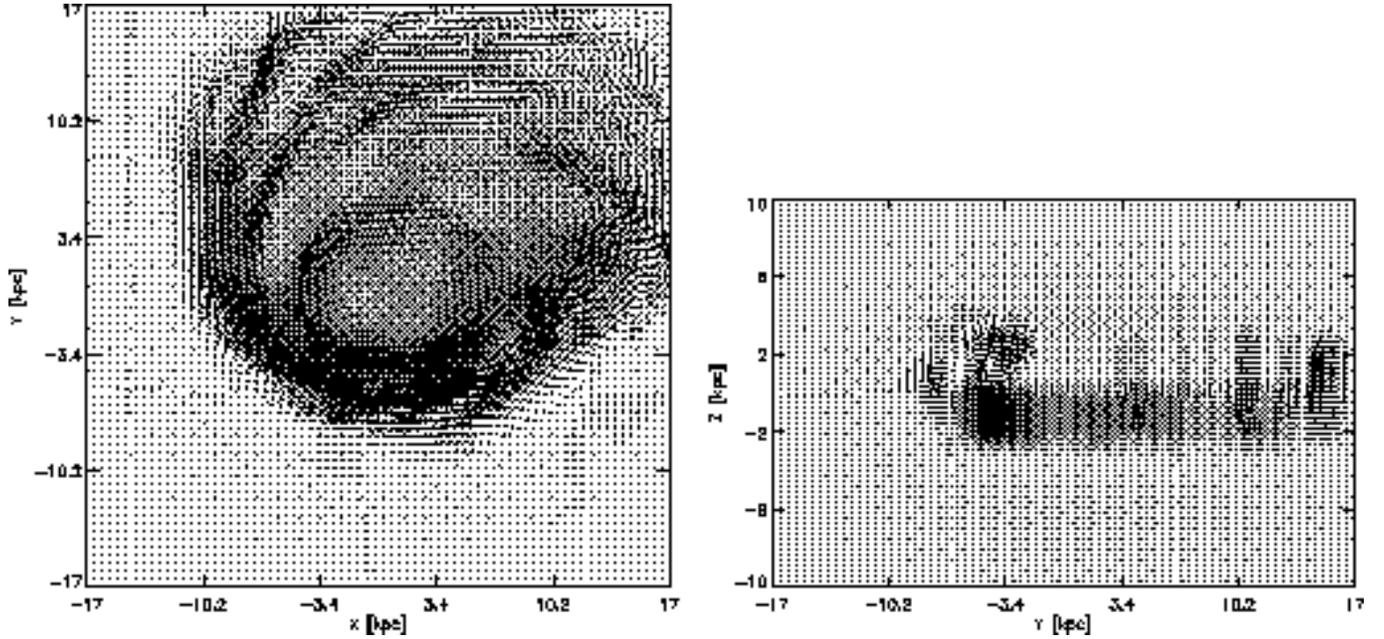}}
\caption{\label{fig1} Magnetic field vectors and gas density in the log scale
in the midplane  at $t \sim$450~Myr (left). 
The closest galaxy's closest
 approach to the cluster center is at $t=500$~Myr. 
Magnetic field vectors and gas density in the log scale
in the plane perpendicular to the galactic disk  at
$t \sim$450~Myr (right). The galaxy moves to the south-east in both figures.}
\end{figure*}
\begin{figure*} 
\resizebox{\hsize}{!}{\includegraphics{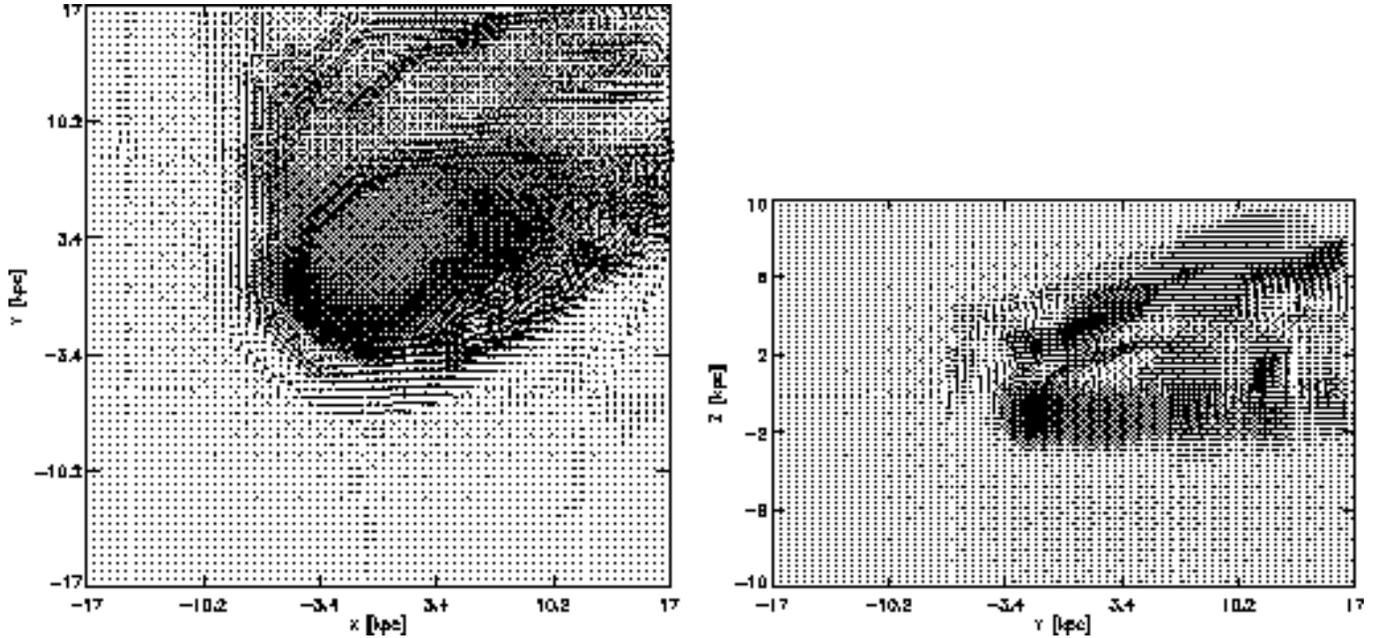}}
\caption{\label{fig2} Magnetic field vectors and gas density in the log scale
in the midplane  at $t \sim$550~Myr (left). 
Magnetic field vectors and gas density n the log scale
in the plane perpendicular to the galactic disk  at
$t \sim$550~Myr (right).
The galaxy moves to the south-east in both figures.}
\end{figure*}
\begin{figure*}
\resizebox{\hsize}{!}{\includegraphics{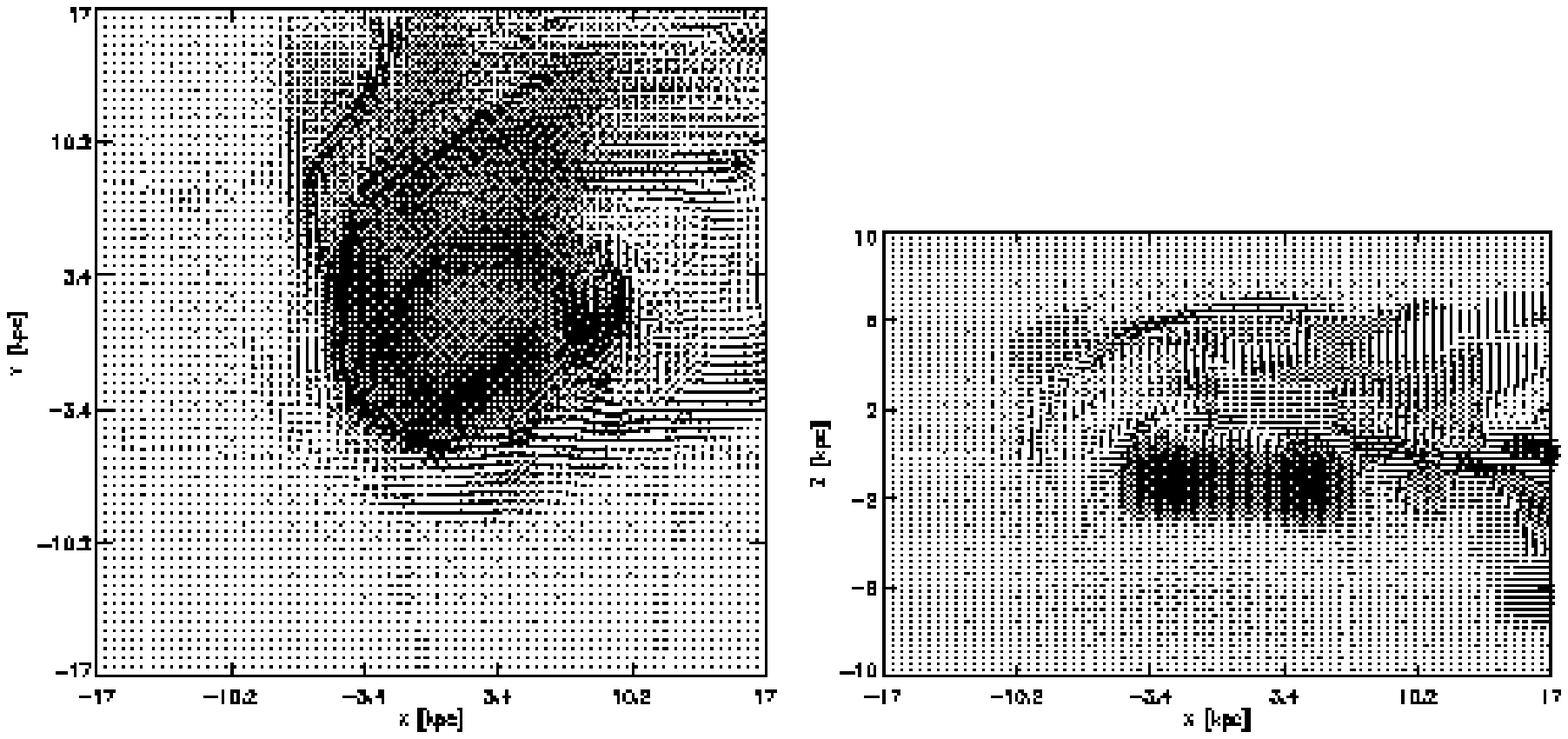}}
\caption{\label{fig3} Magnetic field vectors and gas density in the log scale
in the midplane  at $t \sim$650~Myr (left)  and
in the plane perpendicular to the galactic disk  at
$t \sim$650~Myr (right). The galaxy moves to the south-east in both figures.}
\end{figure*}
All calculated models with  different diffusion coefficient result
in qualitatively similar magnetic field structures. Therefore, we
discuss only case A (uniform diffusion) as the simplest one.
The effects of the ICM ram pressure on the ISM of a galaxy 
moving rapidly through the ambient ICM can be divided into two main
phases: the period of cloud pushing/stripping 
called also the compression phase and of cloud re-accretion 
(see Vollmer et al. 2001).
Both of these processes influence significantly the magnetic field morphology.
The stripping phase takes place roughly between
300 and 600~Myr. Re-accretion happens mainly between 700 and 1000~Myr.

\begin{figure*} 
\resizebox{\hsize}{!}{\includegraphics{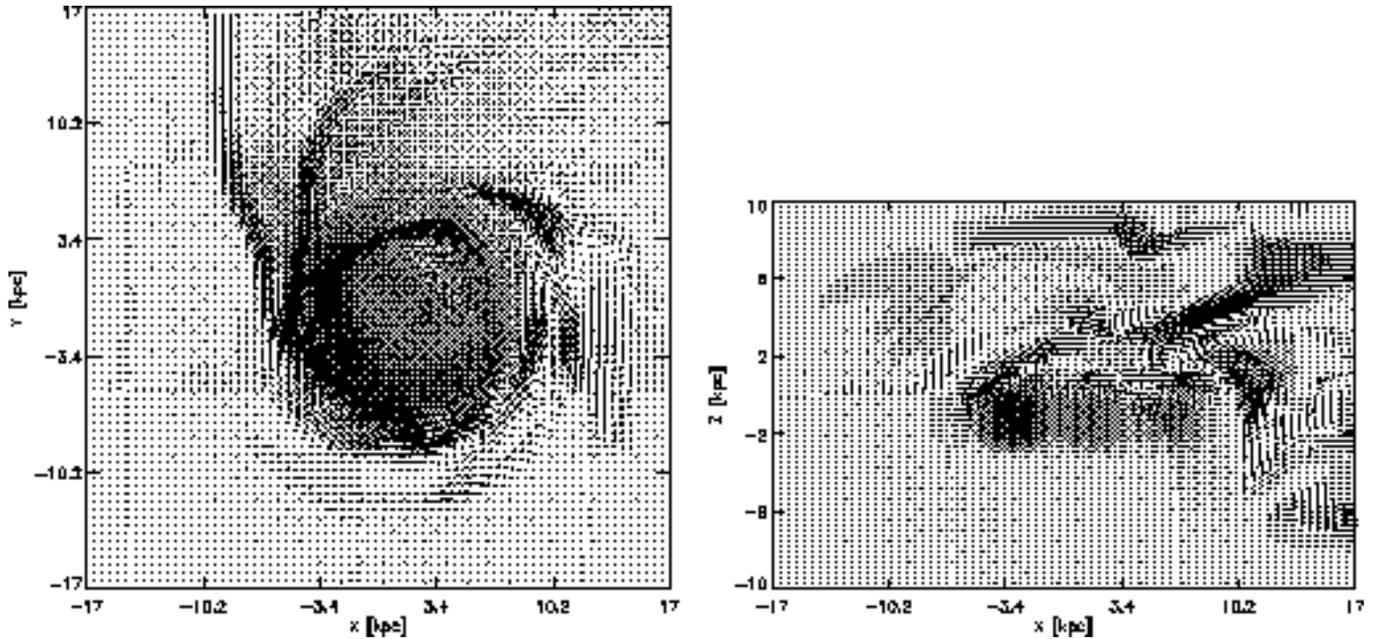}}
\caption{\label{fig4} Magnetic field vectors and gas density in the log scale
in the midplane  at $t \sim$750~Myr (left). 
Magnetic field vectors and gas density in the log scale
in the plane perpendicular to the galactic disk  at
$t \sim$750~Myr (right)}
\end{figure*}

\begin{figure*}
\resizebox{\hsize}{!}{\includegraphics{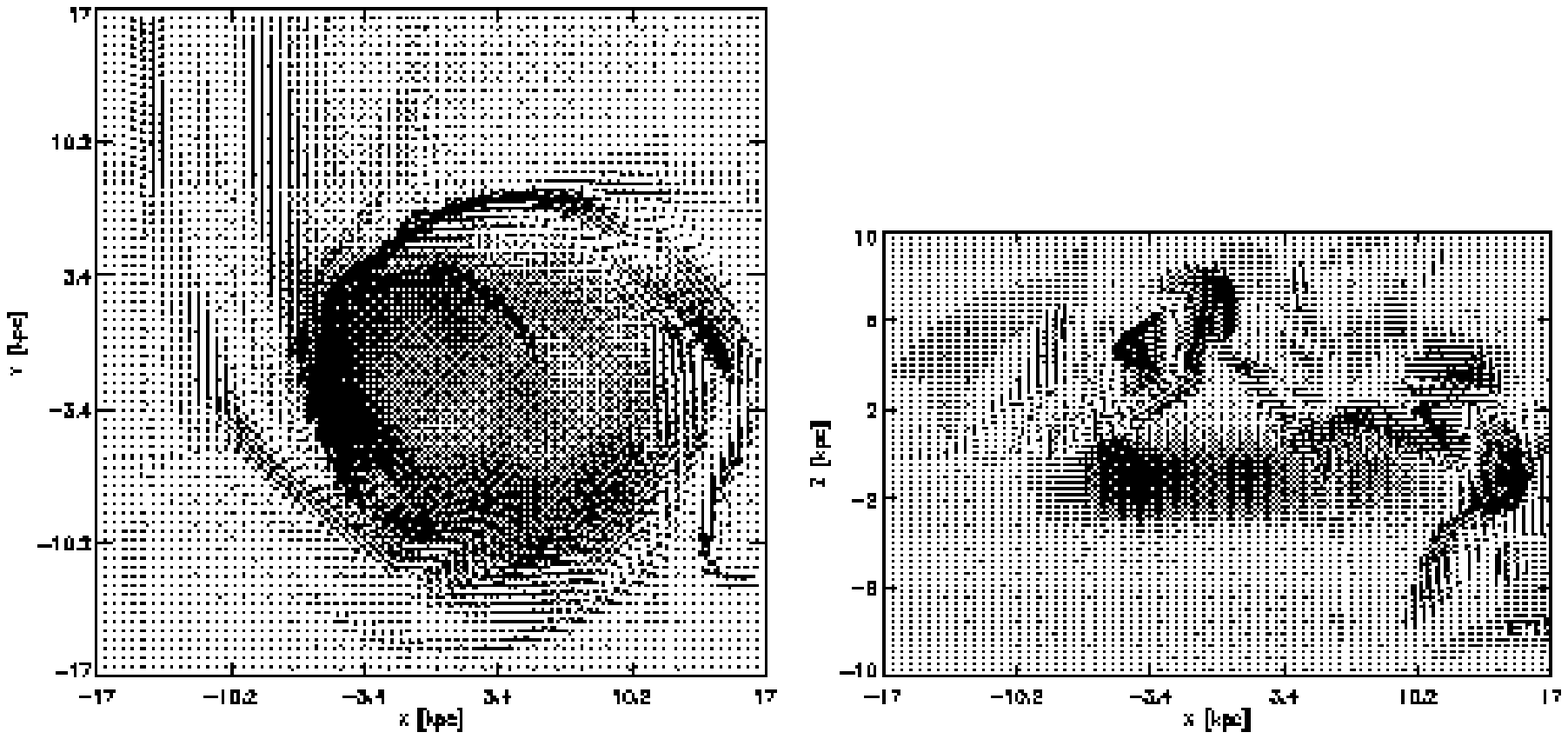}}
\caption{\label{fig5} Magnetic field vectors and gas density in the log scale
in the midplane  at $t \sim$850~Myr (left)  and
in the plane perpendicular to the galactic disk  at
$t \sim$850~Myr (right)}
\end{figure*}

\begin{figure*}
\resizebox{\hsize}{!}{\includegraphics{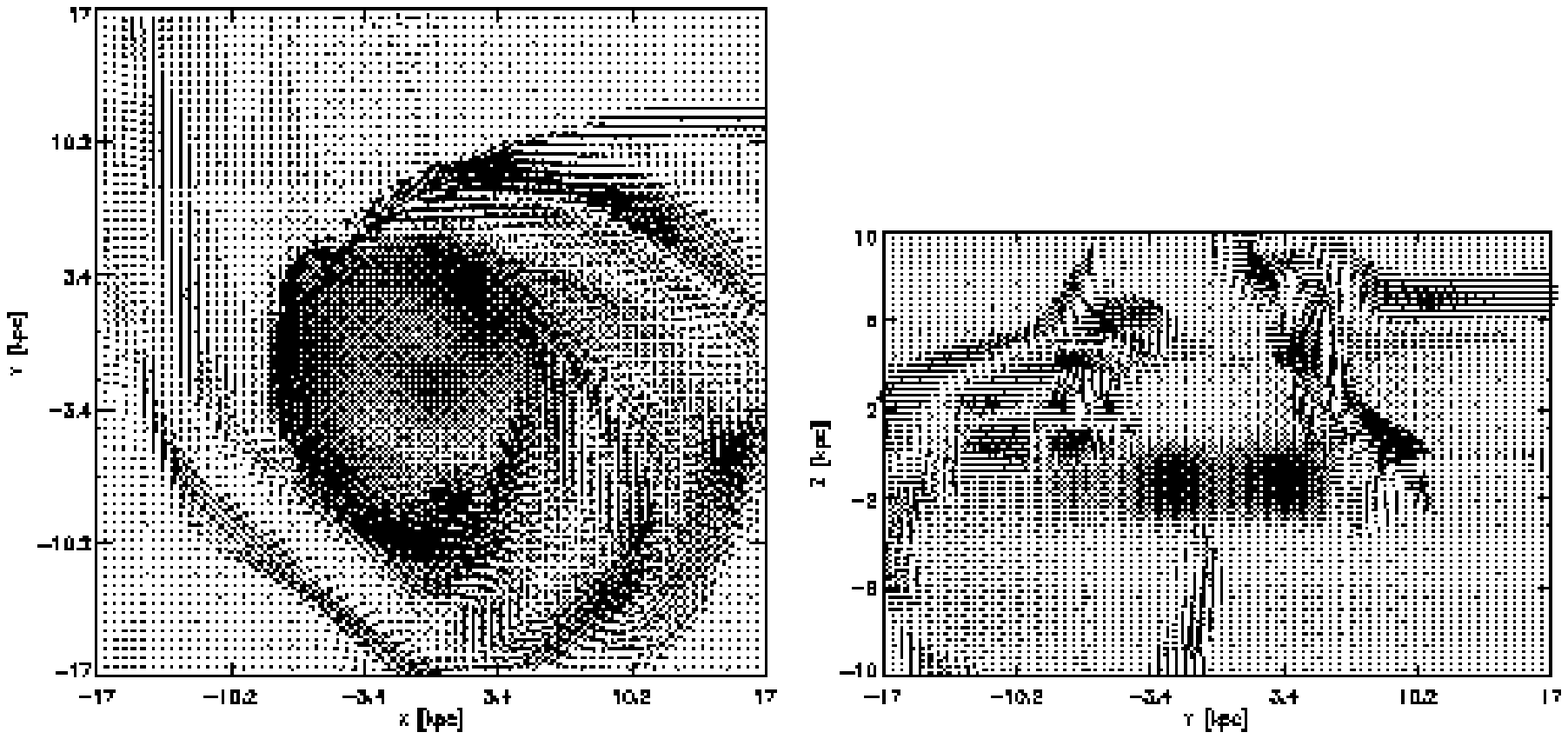}}
\caption{\label{fig6} Magnetic field vectors and gas density in the log scale
in the midplane  at $t \sim$950~Myr (left)  and
in the plane perpendicular to the galactic disk  at
$t \sim$950~Myr (right)}
\end{figure*}

In Figs.1-6 we present snapshots of 6 different timesteps:
$t$=450, 550, 650, 750, 850, and 950~Myr. In each figure the magnetic field
vectors in the midplane of the disk are shown while 
the total gas density is presented 
in a logarithmic scale. The galaxy is viewed face-on 
($X$-$Y$ plane; left panel) or 
edge-on (right panel). In order to emphasize the most interesting
features we show the $X$-$Z$ plane in Fig.~\ref{fig3}--\ref{fig5} and
the $Y$-$Z$ plane in Fig.~\ref{fig1}, \ref{fig2} and  \ref{fig6}.
The maximum lengths of the magnetic field vectors are $B=5 \mu$G
in the $X$-$Y$ plane and $B=2.5 \mu$G in the $X$-$Z$ and 
$Y$-$Z$ planes respectively.
The galaxy moves towards south-east,  which is equivalent to the wind flowing 
from this direction. In the $X$-$Z$ and $Y$-$Z$ plane the wind also comes from the south-east.

\subsubsection{The compression phase \label{sec:comphase}}

Shortly before the ram pressure reaches its maximum ($t$=450~Myr), 
the gas is strongly compressed forming a heavy spiral arm on the windward side 
of the disk (Fig.~\ref{fig1}). The magnetic field is swept together
with the gas  and also forms  a strong maximum in the direction of 
the wind (south-east). Since the compressed magnetic field rotates counter-clockwise
together with the gas (frozen-in field),
its maximum is located in the south-west of the galaxy.
We observe a shift between the maximum of the magnetic field and that of the gas.
Because there is no ISM located further out to the south--east than the gas maximum,
the offset magnetic field can not be associated with the ISM.
Three physical mechanisms can in principle create
such an offset: (i) the pile--up of magnetic field associated with the ICM,
(ii) turbulent diffusion, and (iii) shear amplification between the 
rotating galaxy and the intra-cluster medium.
Since there is no explicit ICM and thus no magnetic field associated with the ICM 
included in the model and diffusion is much too slow to play a role, mechanisms (i)
and (ii) are excluded. In order to estimate the growth of the magnetic field
due to shear, we compare it to the growth due to gas compression:
\begin{equation}
\frac{\partial B_{\phi}}{\partial t}=-B_{\phi} \frac{\partial v_{\rm r}}{\partial r}\ .
\end{equation} 
The growth of the magnetic field due to shear is:
\begin{equation}
\frac{\partial B_{\phi}}{\partial t}=B_{\rm r} r \frac{\partial \Omega}{\partial r}\ ,
\end{equation}
where $B_{\phi}$, $B_{\rm r}$ are the components of the magnetic field in
azimuthal, radial direction, $v_{\rm r}$ is the velocity component in radial
direction, $\Omega$ is the angular velocity, and $r$ is the galactocentric distance.
The derivative $\frac{\partial v_{\rm r}}{\partial r}$ can be estimated by
$\frac{\Delta v_{\rm r}}{\Delta r} \sim$~40~km\,s$^{-1}$/5~kpc.
Since the wind accelerates the outer gas particles $\frac{\partial \Omega}{\partial r}$
is reduced, thus' $r\frac{\partial \Omega}{\partial r} < \Omega <$~150~km\,s$^{-1}$/5~kpc.
In the south-east the magnetic field is mainly azimuthal $B_{\phi} \gg B_{\rm r}$.
Thus, the amplification of the magnetic field due to gas compression exceeds
that due to shear. We conclude that
the shift between the gas and magnetic field maxima in the south-east 
is a pure numerical artifact.
It is due to the use of a density dependent smoothing length together with a 
fixed minimum length.
The smoothing length at the edge of the galaxy's gas distribution is small,
whereas slightly outside, it becomes much larger. The offset thus reflects
the gradient of the smoothing length across the edge of the gas distribution.

In the north-western part of the disk the visible distribution of magnetic 
field structures is caused by stripping effects.
In the western part of the outer disk the magnetic field forms small waves
and changes its direction locally at an almost right angle.
In the $Z$ direction (Fig.~\ref{fig1} right panel) an asymmetry in the 
distribution of magnetic field vectors can be observed in the compression 
region. Since the wind direction has a radial component and a component in the $Z$ 
direction, the magnetic field begins to be pushed radially inwards
and towards positive $Z$.

At the timestep of 550~Myr (Fig.~\ref{fig2}) the gas has formed a prominent 
accelerated arm in the west ($X$=10~kpc, $Y$=3~kpc). 
The magnetic field enhanced due to compression effects
follows the rotation of the gas and moves counter-clockwise.
In addition, being pushed by the wind it moves inwards.  
Its maximum strength is now located west ($X$=10~kpc, $Y$=0~kpc)
of the galaxy closer to its center than for $t$=450~Myr.
The magnetic maximum is offset from the gaseous one, again due
to the gradient of the smoothing length across the edge of the gas distribution. 
The magnetic spiral arms in the north as well as the waving
structure in the west are more pronounced than at $t$=450~Myr.
In the $Z$ direction the compressed magnetic field of the south-eastern part
of the galaxy is swept together with gas by few kpc above the disk plane
(much higher than the initial magnetic field; Fig.~\ref{fig1}).


At the time $t$=650~Myr (Fig.~\ref{fig3}) the galaxy slows down 
on its orbit away from the cluster core. 
The accelerated arm has moved to the 
north of the galaxy and has a complex three dimensional structure. 
The gas maxima start to be redistributed
and wind-compression features diminish their
strength. However, at the south-east disk edge magnetic maxima are still present
(Fig.~\ref{fig3} left).
The northern part of the outer disk is filled with strong magnetic
arms stretching away from the galaxy. 
At the south--west side of the disk ($X$=10~kpc, $Y$=-3~kpc) a counter-rotating 
gaseous arm forms.
The visible maximum of the magnetic strength there is connected with the falling gas.
In the $X$-$Z$ plane the magnetic field extends
to even more positive $Z$ than for $t$=550~Myr, whereas 
the magnetic field at the location, where the counter-rotating arm forms,
extends to negative $Z$.

\subsubsection{The phase of re-accretion}

The period of re-accretion begins at $t \sim 700$~Myr. 
The galaxy moves slowly through outer regions of the cluster.
The wind has ceased and a large number of gas clouds 
starts to fall back onto the galactic disk producing 
very violent gas fountains. The magnetic field is taken up and down
together with the gas and forms poloidal loops extending high in the halo
(even to 7~kpc, see Fig.~\ref{fig5} right panel). 

At $t$=750~Myr (Fig.~\ref{fig4}) the accelerated arm, which falls back onto 
the galactic disk in the north-east, produces a distribution of
maximum gas density similar to that produced by the wind (Fig.~\ref{fig1}).
The counter-rotating arm, which has a very small mass, has no visible influence 
on the gas distribution, but generates magnetic field at negative $Z$ 
that are not visible in Fig.~\ref{fig4}, because it is not located
in the $X$--$Z$ plane.
Due to the strong shear motions at the trailing edge of the back-falling, 
accelerated arm, the magnetic field is enhanced there. It extends towards
positive $Z$.
A second maximum of the magnetic field is visible in the south-west
where an expanding gas shell starts to form.

At $t$=850~Myr (Fig.~\ref{fig5}) the ridge of the enhanced gas density has moved
to the north-east, in a counter-rotating sense. The maximum of the magnetic field
follows this ridge. Additionally, a strong magnetic arm is formed to the north,
whose radial distance from the galactic center increases to the west.
For the first time a weak magnetic arm is visible  east of the galaxy.
In the south-west ($X$=5~kpc, $Y$=-10~kpc) 
the expanding gas shell is now clearly visible.
The magnetic field structure in the $Z$ direction  became very complex.
Several magnetic loops with heights up to several kpc above the disk plane
can be observed in Fig.~\ref{fig5} (right panel).

At the last presented timestep ($t$=950~Myr, Fig.~\ref{fig6}) we observe the 
beginning of a second expanding gas shell south of the galaxy center,
which coincides with a strong ridge of magnetic field. Thus, the beginning 
of the expansion of a gas shell is accompanied by a strong magnetic arm.
In the western part we observe strong open spiral arms. 
The magnetic field structure in the $Z$ direction shows now 
several loops extending up to 10~kpc in both direction of the disk plane.

\subsection{\label{sec:energy} Evolution of the total magnetic energy}

\begin{figure}
\resizebox{\hsize}{!}{\includegraphics{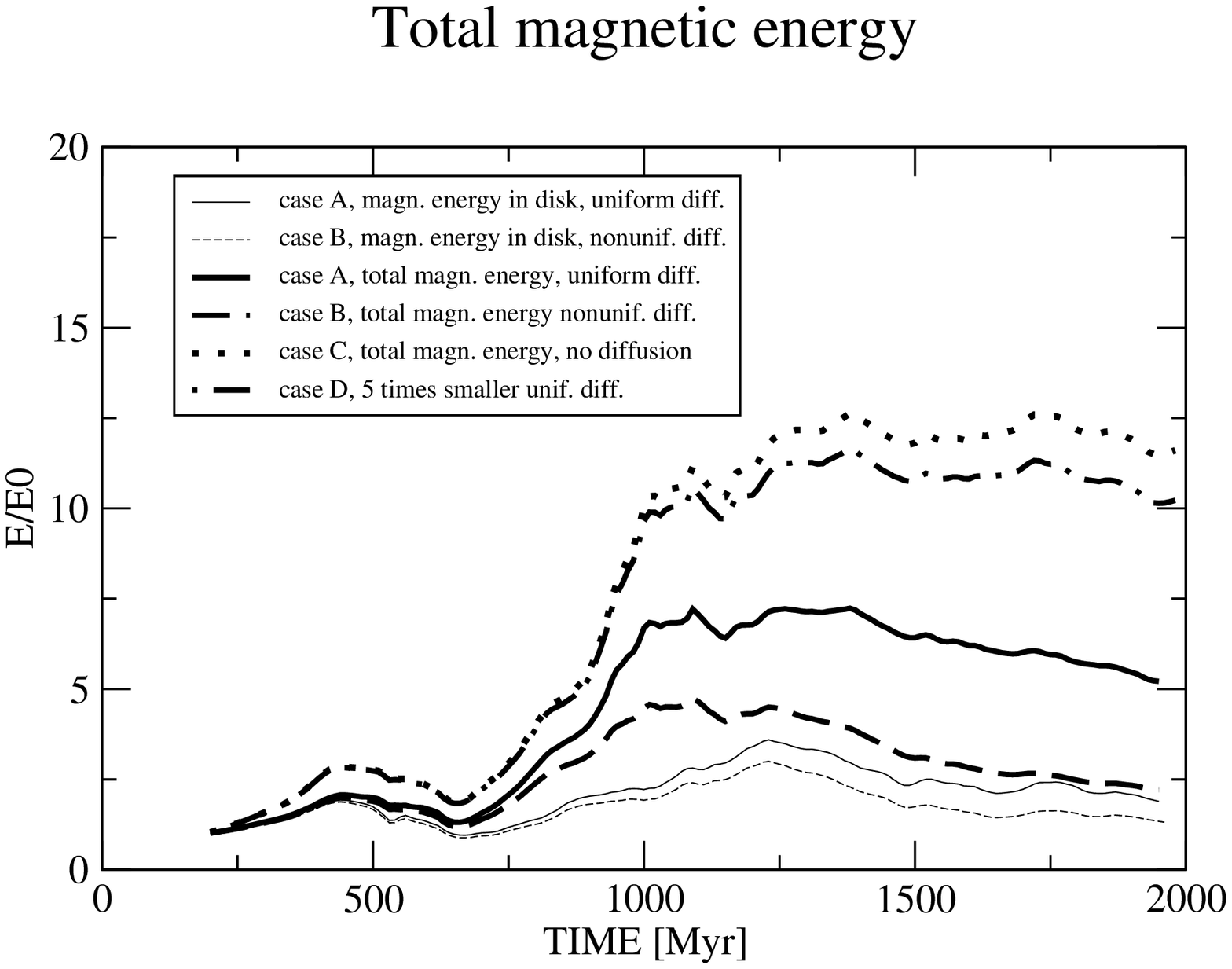}}
\caption{\label{ene} The evolution of the total magnetic energy normalized 
by its initial value as a function of time.
Thin solid line: within a thick disk, uniform diffusion (case A).
Thin dashed line: within a thick disk, non-uniform diffusion (case B).
Thick solid line: within the whole computational box, uniform diffusion (case A).
Thick dashed line: within the whole computational box, non-uniform diffusion (case B).
Thick dotted line: within the whole computational box, without diffusion (case C).
Thick dash-dotted line: within the whole computational box, 5 times lower diffusion than case A (case D).
}
\end{figure}

The evolution of the total magnetic energy normalized to its initial value
is presented in Fig.~\ref{ene} for our 4 experiments: with a uniformly 
distributed magnetic diffusion coefficient (case A),
with a diffusion coefficient growing into the halo (case B),
without a physical diffusion
(case C), and with a diffusion coefficient 5 times smaller than its basic value 
(case D).
We have calculated the magnetic energy (i) in the whole box 
(34~kp$\times$34~kpc$\times$20~kpc) and (ii) in a thick disk centered on the 
galaxy center with a disk height of 1.8~kpc.

For cases A and B the total magnetic energy $E$ increases from its
initial value $E_{0}$ and reaches its maximum $E=2\,E_{0}$ at $t \sim$430~Myr.
In the cases of the reduced physical diffusion (case C)
and no physical diffusion (case D) the total magnetic energy rises
to 3 times its initial value. 
The ram pressure compresses the gas at the windward side and gives 
rise to a radial component of the velocity field there.
This enhances the magnetic field via the $(\vec{v}\times\vec{B})$ term
in the induction equation (Eq.~\ref{eq:inductioneq}).
At the same time the wind expels the magnetic field
out of the galaxy and out of the box of calculation.
>From $t$=500~Myr to 700~Myr the sweeping of the magnetic field
becomes more important than the generation of magnetic field 
due to compression.
The magnetic energy in the disk and in the whole galaxy even decreases  slightly
below its initial value. 

The process of re-accretion, which starts at about 650~Myr, causes 
a fast growth of the total magnetic energy in the whole galaxy from 
$E \sim 4.5\,E_{0}$ (case B, long dashed line) to $E \sim 7.5\,E_{0}$ 
(experiment A, dashed line). Within the thick disk the magnetic 
energy increases to $E \sim 3\,E_{0}$ and $E \sim 3.5\,E_{0}$,
respectively. Moreover, the increase of the magnetic energy within the
disk is slower than in the whole galaxy. In the re-accretion phase the
magnetic energy within the whole box reaches a maximum for case A/B at 
$t \sim 1050$~Myr. The magnetic energy within the disk reaches its maximum
$\sim$150~Myr later. We observe a clear decline of the magnetic energy within the 
disk and the whole galaxy after the maximum for case A and B.

The experiments with a reduced physical diffusion (case D) 
and without physical diffusion (case C) lead to a higher increase
of the total magnetic energy than our basic two experiments A and B.
The energy grows 10 times its initial value during the re-accretion phase.
During the last $\sim$700~Myr the total magnetic energy stays almost
constant for case D and decreases only very slowly for case C.

The influence of the physical diffusion can be directly observed in
comparing case A with cases C/D (Fig.~\ref{ene}). Whereas the graphs for 
case C and D are very close to each other, those of case A (for the whole galaxy) 
are situated well below. Comparing cases A, C, and D for 
1250~Myr$< t <$1350~Myr one can derive a roughly linear correlation 
between total magnetic energy and the diffusion coefficient.
Thus, the physical diffusion clearly dominates over the numerical diffusion.

The linearly growing diffusion coefficient also has a non negligible influence on
the evolution of the magnetic energy. During the re-accretion phase  
its maximum of $E/E_{0}$ is decreased by a factor $\sim$1.4 with respect to 
$E/E_{0}$ for the uniform diffusion coefficient. For later
timesteps this factor decreases to a value of 2.3 at the end of the simulations.
These factors are slightly lower for the magnetic energy computed only 
within the disk.

\subsection{The influence of the cloud--cloud collision rate}

\begin{figure}
\resizebox{\hsize}{!}{\includegraphics{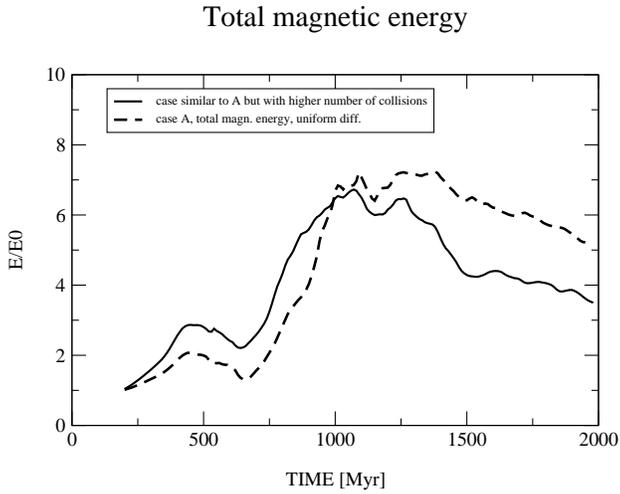}}
\caption{\label{ene4} The evolution of the total magnetic energy normalized 
by its initial value as a function of time.
Dashed line:  uniform diffusion (case A), low cloud--cloud collision rate.
Solid line:  uniform diffusion, 6 times higher cloud--cloud collision rate.
}
\end{figure}

In order to investigate the influence of the cloud--cloud collision rate
on the evolution of the total magnetic energy, we have increased the radii
of the clouds by a factor 6. This leads to an increase of the
number of collisions per unit time of a factor 36. The evolution of the total magnetic
energy  for the simulations with the high and the low
cloud--cloud collision rate is shown in Fig.~\ref{ene4}.
In the case of a high cloud--cloud collision rate (solid line) the magnetic field
energy grows faster and to higher values until $t$=450~Myr. 
This is due to formation of spiral structures within the disk, which is enhanced
by a higher cloud--cloud collision rate. For $t>450$~Myr the magnetic energy evolves
in the same way as that of the simulation with the low cloud--cloud collision rate, but
with an approximately constant offset. At $t=1000$~Myr both curves reach the same
maximum. For $t>1000$~Myr the total energy drops  faster in the case of the
high cloud--cloud collision rate. Within 500~Myr it decreases by 
less than half of its
maximum value. This  decrease is caused by the higher damping of vertical and
shear motions due to frequent cloud--cloud collisions.

\subsection{Polarization maps \label{sec:polmaps}}

In order to analyze the general properties of the magnetic field we  
use simulated maps of the polarized intensity (PI), 
constructed on the basis of our resulting 
magnetic fields. This process yields the information which can be directly
compared with polarized radio continuum observations.  
We performed it in the way similar to that described in our previous works 
(e.g. Otmianowska-Mazur et al. 2000, Otmianowska-Mazur et al. 2002).
The Stokes parameters I, Q and U are integrated along the line of 
sight and convolved  with an assumed beam of 10~$''$. 
We then computed maps of polarized
intensity and the angles of polarization B-vectors.

\begin{figure*}
\resizebox{14cm}{!}{\includegraphics{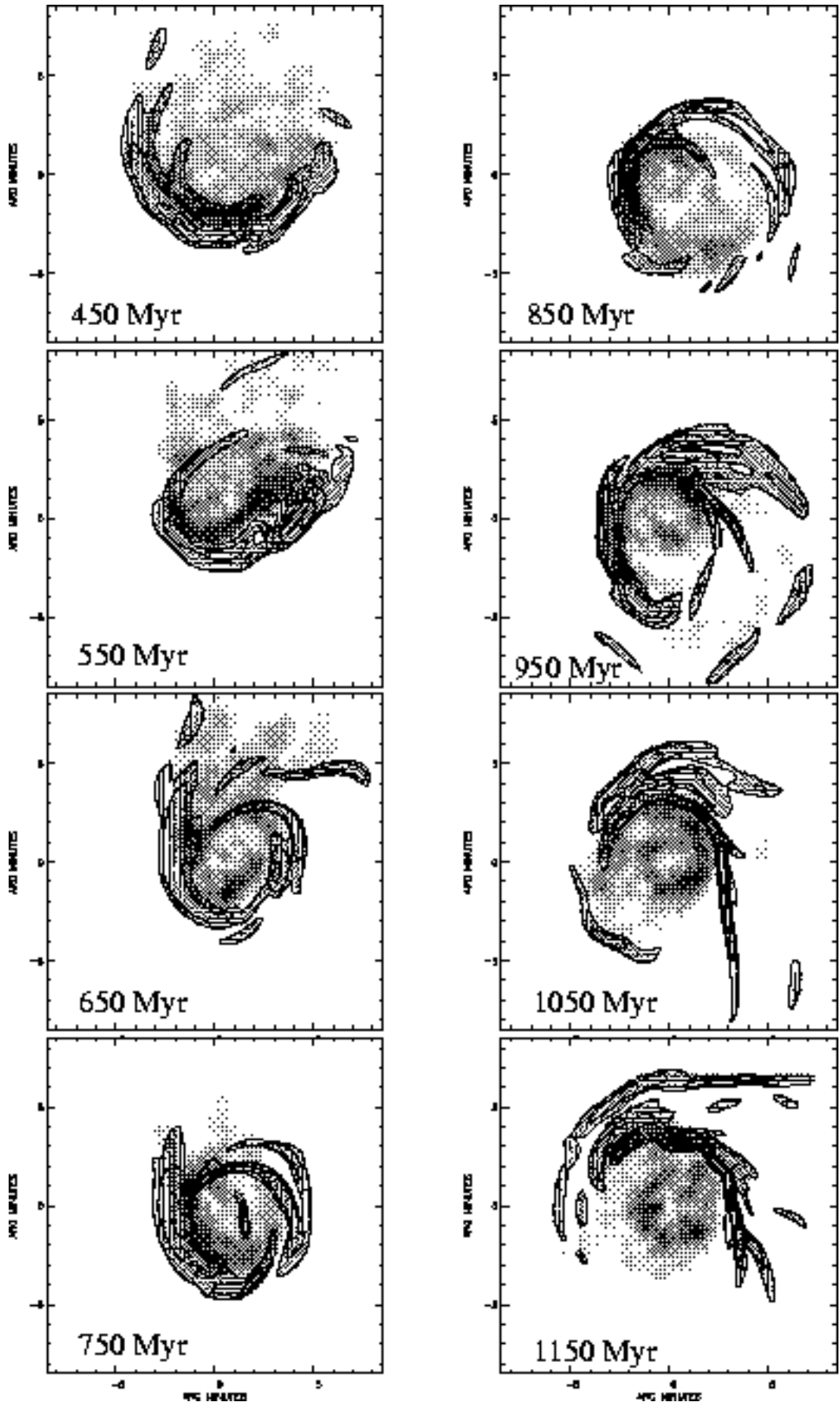}}
\caption{\label{PI} 
Evolution of the polarized intensity at the chosen time steps.
Contours: polarized intensity in logarithmic scale. The magnetic field
vectors are superimposed onto the gas surface density in logarithmic scale. 
The timestep of snapshots is 
written in the lower left corner of each snapshot. The assumed beamsize is 10$''$.}
\end{figure*}

At selected time steps (see Fig.~\ref{PI})
of our model we construct maps of the radio polarization 
intensity (PI) with the inclination of $0^\circ$ (face-on)  
to the sky plane. We use the density distribution of relativistic electrons
that has the form 
\begin{equation}
n_{\rm rel} \propto \exp(-(R/R_{0})^{2}) \exp(-(z/z_{0})^{2})\ ,
\end{equation}
where $R$ is the radial distance to the galaxy center and $z$ is the vertical
distance from the midplane of the disk. We set $R_{0}$=10~kpc
and $z_{0}$=1~kpc. The distance of the galaxy is assumed to be
$\sim$34~Mpc. Thus, one arcminute roughly corresponds to one kpc. 
The beamsize is 10$''$.
The PI maps of model A with high and low cloud--cloud collision rates are very similar.
We show the polarized intensity together with the polarization B vectors
and the gas surface density in Fig.~\ref{PI} for model A with the uniformly distributed diffusion and the low cloud--cloud collision rate.

We will focus here only on the brightest PI emission features that arise from the 
ICM--ISM interaction. They can directly be compared with radio continuum 
observations of cluster galaxies.
At $t$=450~Myr the gas is compressed in the south-east
leading to an excess of density in the gas and a bright 
ridge of polarized emission. This ridge is symmetric and has the 
form of a half-ring. The offset in our model snapshot is again
a numerical artifact due to the gradient of the smoothing length at the edge of 
the galaxy's gas distribution (cf. Sect.~\ref{sec:comphase}).

Shortly after the galaxy's nearest
approach to the cluster center ($t$=550~Myr) 
the accelerated arm is formed in the west.
It is again accompanied by a ridge of the strong polarized emission, which has its 
maximum in the west of the galaxy center. 
Again, the PI emission maximum is shifted outward in comparison with the
maximum of the gas surface density. 

At $t$=650~Myr the re-accretion phase 
begins and the gas falls back onto the disk in the north-east of the galaxy.
In the east, where the gas hits the galactic disk, a strong polarized emission region 
forms. We observe also
ridges of PI at the interface between the infalling and galactic gas
in the north.
In addition, one can still observe a polarized intensity ridge in the west,
that represents the maximum PI emission. 
Both features form an incomplete ring structure. 

This ring structure becomes more complete and symmetric at $t$=750~Myr
with the maximum of PI located in the east of the disk being again slightly shifted
outward with respect to the maximum of the gas surface density.
This shift is real, because it is caused by backfalling gas that hits the disk
in this region. 

At $t$=850~Myr the gaseous clouds
re-accrete form the north and hit the disk in the north-east. The ring
structure in PI that is observed at $t$=750~Myr has now expanded. It becomes 
weaker where the expanding shell forms. The PI maximum is located in the
north-east. A gas shell starts to expand to the south-west partly
accompanied by the polarized intensity.

At the timestep of $t$=950~Myr a large and extended
region of PI forms in the north-west where the gas of the expanding shell
flows back onto the galactic disk. Most of the flux in polarized emission is 
located in the north of the galaxy.
The most southern PI maximum is located where the expanding shell
is anchored.

This north--western region of the extended polarized emission moves in the direction
of rotation 
and appears  in the north of the galaxy at $t$=1050~Myr.
One can still observe a residual ridge of PI in the south-east and a new narrow
feature of PI running north-south.

For the evolution of the PI, one can again distinguish two different evolutionary
phases: (i) the compression phase and (ii) the re-accretion phase.
While the ridges of maximum PI are mainly caused by the
compression for $t < $600~Myr, they are amplified
due to shear via re-accretion for $t > $600~Myr. 
Whereas the maxima
of polarized emission should coincide with the maxima of the gas surface density
in the compression phase, they can be located outside in the re-accretion phase.

\section{Discussion \label{sec:discussion}}

Our numerical experiment clearly indicates that  the
magnetic field is amplified in the whole disk and even outside the galaxy
during the whole period of the evolution.
There are several physical processes responsible for this fact. 
In the first phase, when the galaxy moves through the ICM, the  mechanism
of a strong compression takes place at the windward side of the galaxy. The azimuthal
component of the magnetic field is easily amplified in this process, but also a radial
component is created due to wavy gas motions (see Fig.~\ref{fig2} left panel). 
At later times of the galaxy evolution this radial
field is transformed by the differential rotation into an azimuthal field
(similarly to the MHD dynamo). This represents the most effective
process of magnetic field growth. 
The re-accretion phase is much more complicated than the compression phase. 
The presence of very violent inflows and outflows, shears
connected with them and a differential rotation (resulting in
the reproduction of azimuthal field) gives again possibility of an efficient increase 
of the total magnetic energy.
Inflowing gas moves in a spiral-like manner towards the disk forming
strong magnetic field outside the disk and gaseous maxima at the 
interface with the disk gas.

\begin{figure}
\resizebox{\hsize}{!}{\includegraphics{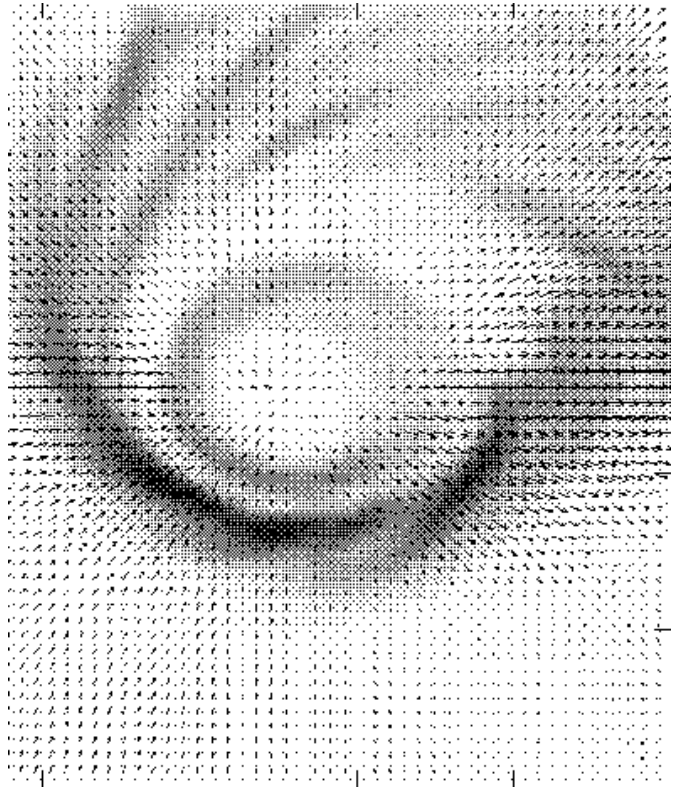}}
\caption{\label{f45vr} The radial component of the velocity field superimposed onto 
the magnetic field strength in the galactic plane at the time step of $t$=450~Myr}
\end{figure}

\begin{figure*}
\resizebox{\hsize}{!}{\includegraphics{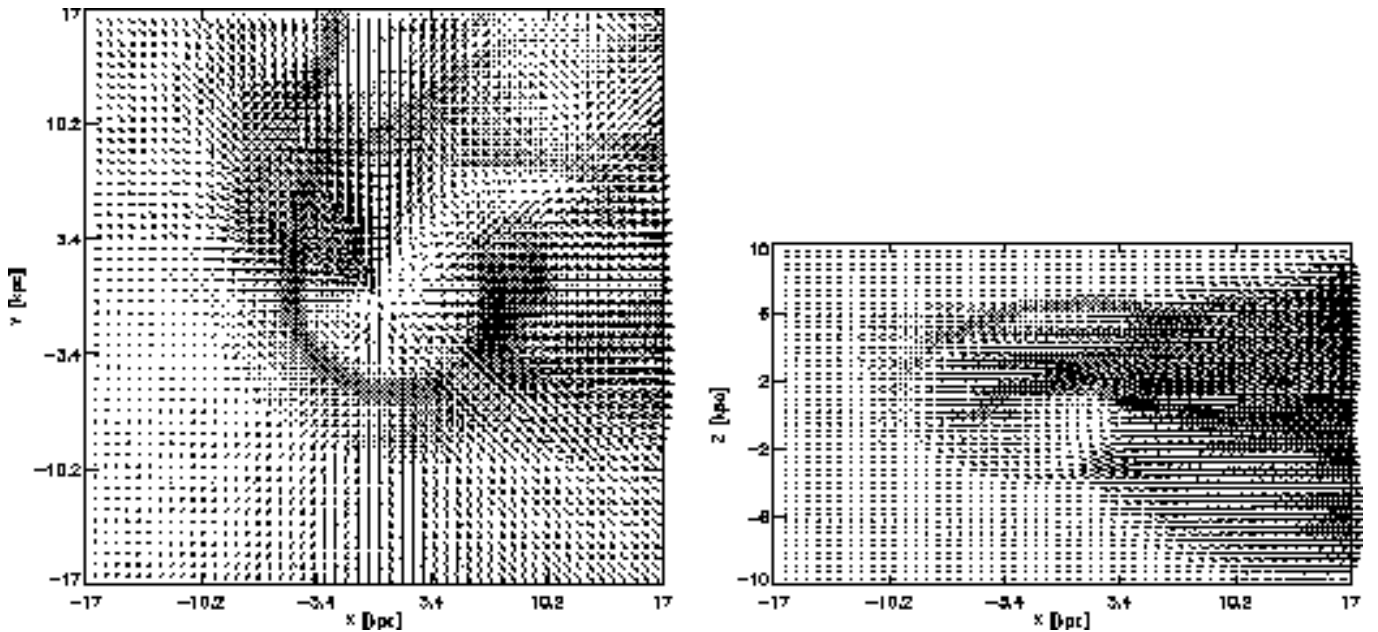}}
\caption{\label{figv65} Velocity vectors superimposed onto 
the magnetic field strength (greyscales) at $t$=650~Myr..
Left panel: the radial velocity component in the galactic plane.
Right panel: the full velocity in the plane perpendicular to the galactic disk.}
\end{figure*}

In order to explain a complicated picture of relations of the velocity
and magnetic field relations we present different velocity components 
as vectors projected onto the magnetic field strength shown as a grey-plot. 
Fig.~\ref{f45vr} shows the galactic plane at the time step of $t$=450~Myr 
(the wind compression phase), while Fig.~\ref{figv65} 
presents two perpendicular cuts of the box at $t$=650~Myr (the re-accretion period).
We choose the same time steps as in Fig.~\ref{fig1} and Fig.~\ref{fig3}.
Our modeled velocity is extrapolated into the whole box in order
to  avoid too strong velocity gradients in our calculations. This is the only
way to avoid an artificial shear at the edges of the galaxy's gas distribution.

The radial velocity (Fig.~\ref{f45vr}) serves as the best indicator of compression 
regions, where the strong gradients of this velocity component are observed.
One of them is visible at the south-east part of the galaxy
and is placed in the direction of the wind causing the formation of 
the magnetic field maximum  in the central part of this quarter
(near the gaseous arm in Fig.~\ref{fig1}, left panel). Above this region in the 
the north-eastern part of the map one can see the radial velocity, which is 
related to  a gas flow from the north. A part of the gas is swept 
from the galaxy to the west. 
 The presented radial velocity component creates a radial magnetic field
component. The differential rotation transforms this radial component into an azimuthal
field and amplifies in this way the magnetic stripes occuring  in this region. 
West of the galaxy center the radial velocity shows strong outflow of gas. 
The magnetic field there is amplified due to the same mechanism: 
strong compression causes the creation of a
radial magnetic field component. The differential rotation transforms it into a toroidal
component amplifying the already existing toroidal magnetic field.
(cf. Fig.~\ref{fig1}, left panel). 

In order to present processes of magnetic field amplification
connected with the re-accretion period,
the left panel of Fig.~\ref{figv65} shows the radial velocity 
component vectors in the plane of the disk at the beginning of the re-accretion phase
($t$=650~Myr).
The strong radial velocity connected with the wind is not longer present
south-east of the galaxy. 
In the south-west the gas is moving outwards. This is the beginning
of the shell expansion. The velocities of the back-falling counter-rotating arm are
not visible, because this arm is not located within the disk plane.
In the north the gas  falls onto the galactic disk causing
the formation of large number of magnetic maxima there.

The right panel of Fig.~\ref{figv65} presents velocity vectors in the $X$-$Z$ 
plane of the modeled box. At the east the magnetic field is  
amplified by the flow. In the lower right quarter of the image the velocity vectors
are almost parallel to the disk, whereas those of the upper right quarter
point towards the disk plane. Such a complicated velocity picture should be studied
with a detailed analysis of the 3D flow, in order to investigate
which mechanism (compression, shear, inflow,
outflow) is responsible for the magnetic field amplification.

The most interesting and new result of our calculations is 
the growth of the total magnetic energy during the phase
of re-accretion. We found
that the total magnetic energy grows up to a few times its initial value 
(cf. Fig.~\ref{ene}) within the thick disk. Three processes are
responsible for this growth: (i) compression of
magnetic field by re-accreting gas, (ii) enhancement of magnetic field due to shearing
motions, and (ii) a new kind of dynamo mechanism:
the re-accreting material has important velocities in the radial 
and the vertical direction. Together with the differential rotation the galaxy's 
velocity field becomes very complex. 
For $t>700$~Myr large gas loops are formed in the halo with heights up to
several kpc. The magnetic field associated with them
has an important poloidal component.
The differential rotation in the disk creates the toroidal component 
of the magnetic field from the poloidal component (see Ruzmaikin et al. 1988). This may 
constitute a mechanism of the magnetic field generation.
Thus, we find a new kind of the dynamo working in cluster galaxies, which have undergone 
a ram pressure event and are now in the phase of re-accretion.

The modeled polarization maps show a characteristic evolution of the 
distribution of the
polarized intensity (PI). This evolution can be divided into two
distinct phases:
\begin{itemize}
\item
the compression phase and  
\item
the re-accretion phase.
\end{itemize}
During the compression phase the maximum of the PI 
should coincide with the maximum of
the gas surface density. This is not necessarily the case during the re-accretion phase.
Thus, the polarized radio continuum emission could represent a tool to discriminate 
between different evolutionary stages of a cluster galaxy that undergoes 
or has undergone an ICM--ISM interaction. Together with the gas distribution and
velocity field the PI distribution gives an important additional 
constraint on dynamical models. We plan to check this hypothesis in our future study.
Observationally two Virgo galaxies that were observed in PI with the VLA at
2.8~cm and 6~cm (NGC4254: Soida et al. 1996 and 
NGC4522: Vollmer et al. in prep.). Both show asymmetric ridges of PI.  
Our future project is to
compare the modeled PI maps to the high-frequency (Faraday rotation-free) 
polarized radio emission observations of Virgo cluster galaxies.

\section{Conclusions \label{sec:conclusions}}

The evolution of the large-scale galactic magnetic field in a cluster galaxy 
undergoing an intracluster -- interstellar
interaction is investigated using 3D numerical simulations.
In order to solve the induction equation (Eq.~\ref{eq:inductioneq})
we use velocity fields resulting from an N-body, sticky particle code 
including ram pressure effects (Vollmer et al. 2001).
 
We considered the general problem of magnetic field evolution in
galaxies interacting with the ICM.  Two experiments are performed 
with different values of magnetic diffusion. The classical $\alpha$-effect 
has not been involved in our computations.  We perform the
simulations of a galaxy, which is 
mildly HI deficient after the stripping event, as a representative model 
of the N-body calculations. This simulation presents the most characteristic features 
of magnetic field evolution during and after an ICM--ISM interaction.   

The modeled magnetic fields are used to construct maps of the distribution of
polarized radio continuum emission, which can be directly compared to observations.

We find that:
\begin{itemize}
\item  
interactions with the intracluster medium 
(i.e. the galaxy motion with respect to the ambient medium) lead to the 
formation of a heavy spiral arm along with a strong magnetic field
concentration, on the disk windward side. On the other side  of the disk the gaseous and 
magnetic arms stretch away from the disk center.
\item  
Due to the gas re-accretion after the stripping event the vertical magnetic fields 
arise, forming fragments of a large-scale poloidal field, resembling those created by 
a dynamo process.
\item  
Due to the ICM--ISM interaction, the total magnetic energy in cluster
galaxies, which are affected by ram pressure, grows without any dynamo action. 
The interaction with the ambient gas may serve as a magnetic field amplification process 
in cluster spirals. 
\item
The distribution of polarized radio continuum emission shows characteristic features
during the galaxy evolution within the cluster.
The comparison of maps of polarized radio continuum emission with our simulations will
help to determine the evolutionary stage of a galaxy that undergoes an ICM--ISM interaction.
It will also serve as an important
additional constraint on dynamical modeling of ICM--ISM interactions.
\end{itemize}

\acknowledgements
The authors express their attitude to Prof. Chantal Balkowski
and Prof. Marek Urbanik for helpful discussions. 
K.O.-M. and B.V. thank Dr. Marian Soida for a detailed discussion of the 
polarization maps. This work was partly supported by a
grant from the Polish Committee for Scientific Research (KBN), grant no.
0249/P03/2001/21. 


\begin{thebibliography}{}

\bibitem{a1} Abadi, M.G., Moore, B., \& Bower, R.G. 1999, \mnras, 308, 947

\bibitem{a2} Balsara, D., Livio, M., \& O'Dea, C.P. 1994, \apj, 437, 83

\bibitem{a3} Bravo-Alfaro H., Cayatte V., van Gorkom J.H., \& Balkowski C. 2000, AJ, 119, 580

\bibitem{a4} Cayatte V., Kotanyi C., Balkowski C, \& van Gorkom J.H.  1994, AJ, 107, 1003 

\bibitem{a5} Chy\.zy K.T., Soida M., Urbanik M., Beck R., 2001,
in Proc. 24th General Assembly of IAU, Manchester 7-18 August 2000, in press.

\bibitem{a6} Combes F., Dupraz C., Casoli F., \& Pagani L. 1988, A\&A, 203, L9

\bibitem[2000]{els00} Elstner D., Otmianowska-Mazur K., v. Linden S., 
Urbanik M.: 2000, A\&A 357, 129

\bibitem{a7} Gaetz, T.J., Salpeter, E.E., \& Shaviv G. 1987, \apj, 316, 530

\bibitem{a8} Otmianowska-Mazur K., von Linden, S., Lesch, H., Skupniewicz, G.,
 1997, A\&A,  323, 56

\bibitem[2000]{otm00} Otmianowska-Mazur K., Chy\.zy K., Soida M.: 2000, A\&A 359, 29

\bibitem{a9} Otmianowska-Mazur K.,  Elstner D., Soida M., Urbanik M., 2002, A\&A, 384, 48

\bibitem[Parker 1979]{Parker79} Parker, E.N.: 1979, {\em Cosmical Magnetic Fields}, Oxford

\bibitem{a10} Ruzmaikin A.A., Shukurov A.M., \& Sokoloff D.D. 1988, Magnetic Fields of Galaxies, Astrophysics and space science library, Vol. 133, Kluwer Academic Publishers

\bibitem{a11} Schulz S. \& Struck C. 2001, MNRAS, 328, 185

\bibitem{a12} Soida M., Urbanik M., \& Beck R. 1996, A\&A, 312, 409

\bibitem{a13} Solanes J., Manrique A., García-G\'omez C., Gonz\'alez-Casado G., Giovanelli R., \& Haynes M.P. 2001, ApJ, 548, 97

\bibitem{a14} Stone J.M., Norman M.L., 1992a, ApJS 80, 791 
\bibitem[1992b]{stone2} Stone, J. M., Norman, M. L., 1992b, ApJS, 80, 791

\bibitem{a15} Takeda, H., Nulsen, P.E.J., \& Fabian, A.C. 1984, \mnras, 208, 261

\bibitem{a16} Tosa, M. 1994, \apj, 426, L81

\bibitem{vollmer} Vollmer B., Cayatte V., Balkowski C., \& Duschl W.J. 2001, ApJ, 561, 708

\end{thebibliography}
\end{document}